\documentclass[preprint,12pt,sort&compress]{elsarticle}
\usepackage{amsfonts}
\usepackage{amsmath}
\usepackage{amsthm}
\usepackage{amssymb}
\usepackage{graphicx}
\usepackage{dcolumn}
\usepackage{bm}
\usepackage{bbding}
\usepackage{mathrsfs}
\usepackage{graphicx}

\biboptions{numbers,sort&compress}

\begin{document}

\begin{frontmatter}
\title{The effect of center-of-mass motion on photon statistics}
\author{Yang Zhang}
\author{Jun Zhang}
\author{Shao-xiong Wu}
\author{Chang-shui Yu\corref{cor1}}
\ead{quaninformation@sina.com}
\cortext[cor1]{Corresponding author. Tel: +86 41184706201}
\address{School of Physics and Optoelectronic Technology, Dalian University of
Technology, Dalian 116024, China}
\begin{abstract}
We analyze the photon statistics of a weakly driven cavity quantum
electrodynamics system and discuss the effects of photon blockade and
photon-induced tunneling by effectively utilizing instead of avoiding the
center-of-mass motion of a two-level atom trapped in the cavity. With the
resonant interaction between atom, photon and phonon, it is shown that the
bunching and anti-bunching of photons can occur with properly driving
frequency. Our study shows the influence of the imperfect cooling of atom
on the blockade and provides an attempt to take advantage of the
center-of-mass motion.
\end{abstract}

\end{frontmatter}

\section{Introduction}

The nonlinear interaction represents a hot research topic in recent years.
It can lead to a variety of intriguing nonlinear optical phenomena which are
ubiquitous in quantum optics, such as the vortex formation, self-focusing,
soliton propagation, optical multi-stability and so on \cite{Boyd, Liew}.
The nonlinear interaction between photons, light or under the influence of
Kerr medium \cite{Optics,Schmidt} has also been widely used to generate
nonclassical field states \cite{A. D,F. L} and in the observation of strict
quantum effects. Cavity quantum electrodynamics (CQED), as the light and
matter meeting interface, is the most straightforward way to produce the
quantum nonlinear dynamics \cite{tian,Werner}. A lot of strict quantum
effects such as optical solitons \cite{33}, quantum phase transitions \cite%
{44, 55}, quantum squeezing \cite{66} and optical switching with single
photon \cite{77} have been demonstrated in CQED systems based on the strong
optical nonlinearity.

As a typical nonlinear quantum optical effect, photon blockade shows that
the system `blocks' the absorption of a second photon with the same energy.
The typical feature of this effect is the photon anti-bunching which is
signaled by a rise of $g^{(2)}(\tau )$ with $\tau $ increasing from 0 to
larger values while $g^{(2)}(0)<g^{(2)}(\tau )$ as discussed in detail in
Ref \cite{Mandel}. Recently, quantum blockade schemes have been reported in
various physical setups such as superconducting circuit \cite{Hoffman,Liu},
cavity arrays \cite{I. Carusotto,Brandao,55,Bose}, quantum dots \cite%
{Winger,pan}, atomic systems \cite{tan,Brandao,Plenio2}, quantum
optomechanical setups \cite{Rabl,Nunnenkamp,87,liao}, confined cavity
polaritons \cite{Verger}, systems with ultrastrong coupling \cite{Leib} and
so on \cite{Lounis,Kuzmich,Mayer}. What is more , the strong bunching
effect, and its special signature called photon-induced tunnelling, which
shows the absorption of the first photon can enhance the absorption of the
subsequent photons are also been studied \cite{xu,coherent}. As to the
photon blockade in a cavity, the fundamental mechanism is the anharmonic
energy-level structure of the light field when the photon-photon interaction
is induced by the nonlinear medium \cite%
{Schmidt,tian,77,Mandel,Nori,yong,becher,aoki,hong,dd} or induced by the
interaction between the trapped multi-level atom and the multi-mode cavity
\cite{Kimble}, or is observed in the detuned Jaynes-Cummings (J-C) model
\cite{coherent,Arka}.

In this paper, we study the photon statistics in the detuned J-C model and
the effect of the atomic center-of-mass (COM) motion on the photon statistics.
Instead of the multi-level atom and the multi-mode cavity, we require that a
single two-level atom trapped in a single mode cavity driven by a weak laser
and the atom oscillates at its origin with trap frequency. Thus, the
photon-photon interaction is induced by the participation of phonons. It is
found that the photon blockade is generated under the resonant interaction
among atom, photon and phonon, and both the bunching and anti-bunching of
photons strongly depend on the atomic motion. In particular, by comparing
the cases with and without atomic center-of-mass motion, we find that the
atomic center-of-mass motion can enrich the phenomenon of photon statistics.
The advantages: on one hand, are to reveal the influence of the atomic
center-of-mass motion on the blockade due to the imperfect cooling and on
the other hand, are to provide an attempt to taking advantage of the
center-of-mass motion of the trapped atom. This paper is organized as
follows. In Sec 2, we analyze the equal-time correlation function of cavity
mode and demonstrate the photon statistics of J-C model; In Sec 3 we
introduce the atomic center-of-mass motion, and investigate the equal-time
correlation two-time correlation function of photons; Finally, the
conclusion is given in Sec 4.

\section{Equal-time correlation in the detuned J-C model}

A two-level system (ion or an atom) couples to the cavity with frequency $%
\omega _{a}$ which is driven by an external optical field. The frequency of
atomic transition from ground state $\left\vert g\right\rangle $ to excited
state $\left\vert e\right\rangle $ with linewidth $\gamma $ is denoted by $%
\omega _{e}$. The related Hamiltonian under the rotating wave approximation
can be written as \cite{jan}
\begin{equation}
H=\Delta a^{\dag }a+\delta \sigma ^{+}\sigma ^{-}+g(a^{\dag }\sigma
^{-}+a\sigma ^{+})+\Omega \left( a^{\dag }+a\right) ,  \label{a}
\end{equation}%
where $\Delta =\omega _{a}-\omega _{L}$ is the laser detuning from the
cavity mode and $\allowbreak \delta =\omega _{e}-\omega _{L}$ is the laser
detuning from the atom, in the n-excitation subspace, the eigenstates and
the eigenvalues can be easily get \cite{jan,scully}
\begin{equation}
\left\vert n+\right\rangle =\sqrt{\frac{1}{2}-\frac{\tilde{\delta}}{2\Delta
^{\prime }}}\left\vert n,g\right\rangle +\sqrt{\frac{1}{2}+\frac{\tilde{%
\delta}}{2\Delta ^{\prime }}}\left\vert n-1,e\right\rangle ,
\end{equation}%
\begin{figure}[tbp]
\centering
\hspace*{-2cm} \includegraphics[width=1\columnwidth,height=2in]{JCnengji.eps}
\includegraphics[width=0.75\columnwidth,height=1.6in]{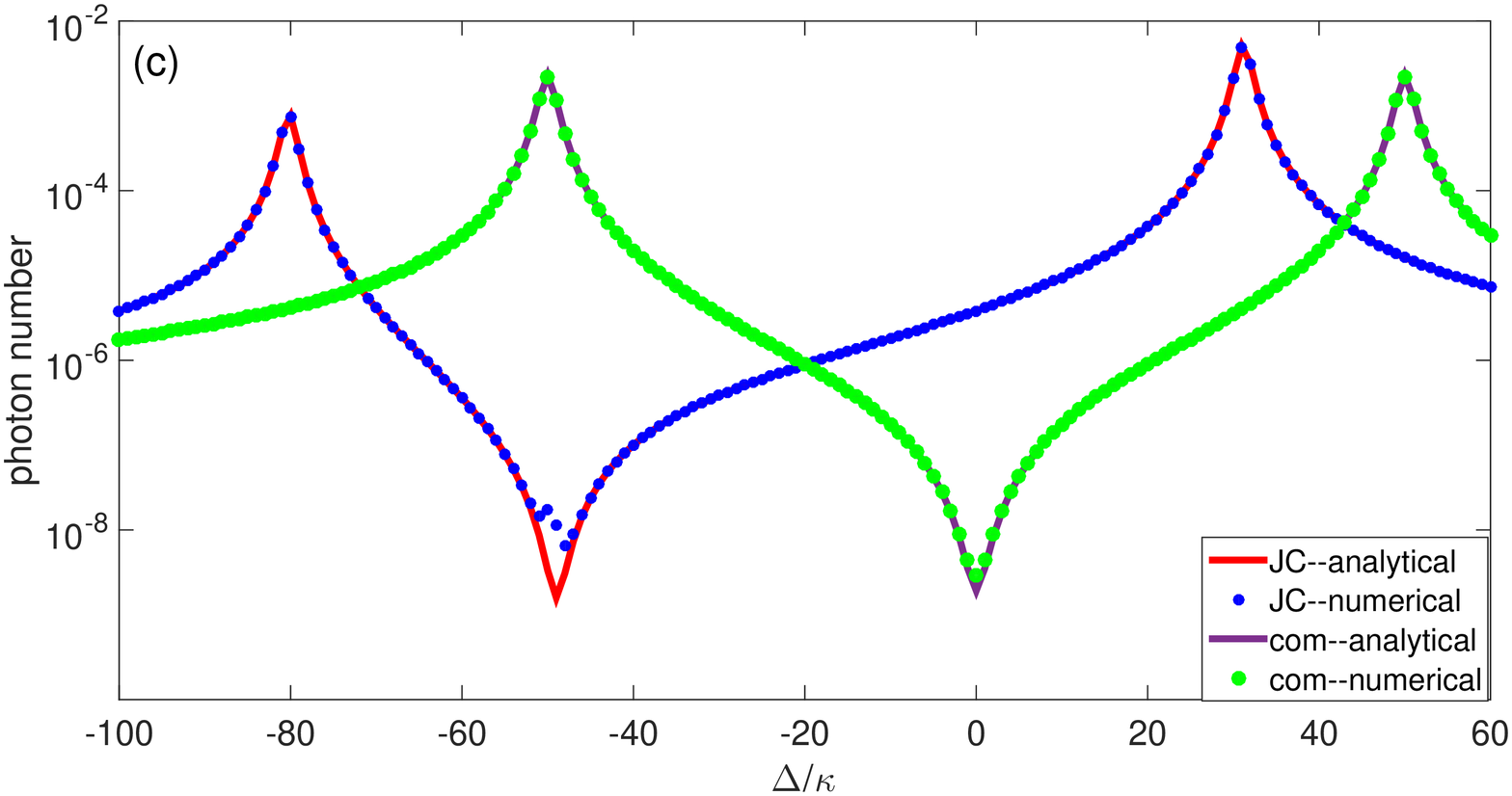} %
\includegraphics[width=0.75\columnwidth,height=1.6in]{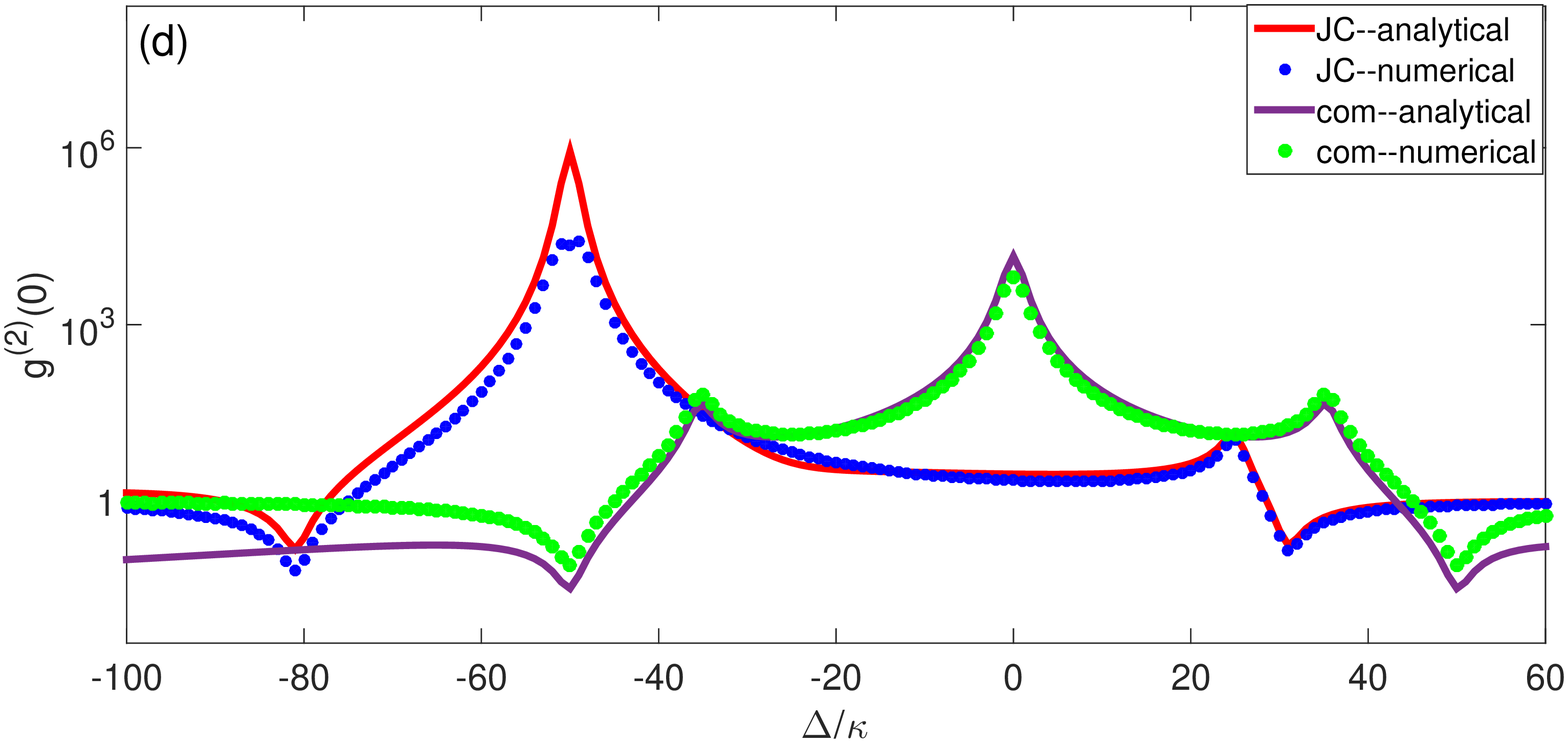}
\caption{(color online).(a) Diagram for the eigensystem of the Hamiltonian
of the coupled cavity-atom system. The anharmonic energy difference between
consecutive manifolds is $\Delta ^{\prime }=\protect\sqrt{4ng^{2}+\tilde{%
\protect\delta}}$ which is not constant. (b) The energy diagram with the
anharmonic spacing energy under the strong coupling conditions (parameters
are chosen as follows). The level shift is caused by the strong coupling.
The blue lines indicate the frequency transitions of single photon process
for the $\left\vert 0\right\rangle \rightarrow $ $\left\vert 1_{\pm
}\right\rangle $ and red line indicate the two-photon process $\left\vert
0\right\rangle \rightarrow $ $\left\vert 2_{\pm }\right\rangle $ . (c) ,(d)
Mean photon numbers of the cavity and the equal-time second-order function
vs. the cavity-laser detuning $\Delta$ for the J-C model and atomic
center-of-mass motion model. The red and purple curves are approximate
analytical solutions of the mean photon number of the J-C model and COM model
respectively, while the blue and green dashed curves stand for numerical
results based on the solution of the quantum master equation Eq. (13) and
Eq. (32). Further explanation of the difference about the two model can be
found in the next section. Here all parameters are dimensionless. The
parameters are $\protect\gamma /\protect\kappa =1,$ $\Gamma /\protect\kappa %
=0.1$, $g/\protect\kappa =50$, $\protect\delta =\Delta+\tilde{\protect\delta}
$, $\Omega /\protect\kappa =0.1$, $\tilde{\protect\delta}/\protect\kappa =50.
$}
\end{figure}
\begin{equation}
\left\vert n-\right\rangle =\sqrt{\frac{1}{2}+\frac{\tilde{\delta}}{2\Delta
^{\prime }}}\left\vert n,g\right\rangle -\sqrt{\frac{1}{2}+\frac{\tilde{%
\delta}}{2\Delta ^{\prime }}}\left\vert n-1,e\right\rangle ,
\end{equation}%
with $\tilde{\delta}=\delta -\Delta $ the detuning in frequency between the
two-level transition system and cavity mode and $\Delta ^{\prime }$=$%
\sqrt{4ng^{2}+\tilde{\delta}}$ the energy difference between manifold
levels. The eigenvalues are $E_{\pm }^{(n)}=(n-\frac{1}{2})\Delta +\frac{1}{2%
}\delta \pm \frac{\Delta ^{\prime }}{2}(n\geqslant 1).$ We can easily
find that the energy splitting has a nonlinear dependence on n (photon
number) and gives rise to the nonlinear optics which contains the photon
blockade effect and tunneling phenomenon. We will show this in the following
part by a strict treatment.

Fig. 1(a) shows the energy diagram of the system. The energy difference
value $\Delta ^{\prime }$ which is a not a constant, this anharmonic
spacing level can affect the photon statistical distribution, such as the
photon blockade and photon-induced tunnelling. To understand the photon
statistical situation, we will focus on the photon correlation that
characterizes the nonclassical photon statistics in the system. Here, we
study the equal-time (namely zero-time-delay) second-order photon-photon
correlation function \cite{scully}:
\begin{equation}
g^{(2)}(0)=\frac{\left\langle a^{\dagger }a^{\dagger }aa\right\rangle }{%
\langle a^{\dagger }a\rangle ^{2}}=\frac{\sum_{n}n(n-1)p_{n}}{%
(\sum_{n}np_{n})^{2}},  \label{nn}
\end{equation}%
where $n=\langle a^{\dagger }a\rangle $ is the intra-cavity photon number of
the cavity mode, $p_{n}$ represents the probability with $n$ photons. In Eq.
(\ref{nn}) the operator is evaluated at the same time. When the photon
anti-bunching occurs, the second-order correlation function should fulfills
the inequality $g^{(2)}(0)\leq 1$ and the limit $g^{(2)}(0)\rightarrow 0$
corresponds to the perfect photon blockade in which two photons never occupy
the cavity at the same time. On the contrary, when $g^{(2)}(0)>1$, it means
photons inside the cavity enhance the resonantly entering probability of
subsequent photons \cite{xuxunwei,lang}.

In order to give an intuitive picture, we take an analytical (but
approximate) method to calculate the second-order correlation function by
employing the wave function amplitude approach. Considering the effects of
the leakage of the cavity $\kappa $, the spontaneous emission $\gamma $ of
the atom, we phenomenologically add the relevant damping contributions to
Eq.~(\ref{a}). Thus the Hamiltonian can be rewritten as $H-i(\kappa
a^{\dagger }a+\gamma \sigma ^{+}\sigma ^{-})$. Since we consider the weak
driving limit, only few photons can be excited in the cavity. So one can
assume that the state of the composite system can be given by \cite%
{fliuds,steady1}
\begin{eqnarray}
\left\vert \Psi \right\rangle &=&A_{0g}\left\vert 0,g\right\rangle
+A_{1g}\left\vert 1,g\right\rangle +A_{0e}\left\vert 0,e\right\rangle  \notag
\\
&&+A_{1e}\left\vert 1,e\right\rangle +A_{2g}\left\vert 2,g\right\rangle .
\label{phi}
\end{eqnarray}%
Since the dynamics of the system is governed by Schr\"{o}dinger equation,
using the Hamiltonian and the state $\left\vert \Psi \right\rangle $ , we
can arrive at the equations about the amplitudes in Eq. (\ref{phi}) as
follows.
\begin{subequations}
\begin{equation}
\dot{A}_{1g}=-(\kappa +i\Delta )A_{1g}-i\Omega A_{0g}-igA_{0e}-\sqrt{2}%
i\Omega A_{2g},  \label{steady}
\end{equation}%
\begin{equation}
\dot{A}_{01}=-(\gamma +i\delta )A_{01}-igA_{10}-i\Omega A_{11},
\end{equation}%
\begin{equation}
\dot{A}_{2g}=-2(\kappa +i\Delta )A_{2g}-\sqrt{2}igA_{1e},
\end{equation}%
\begin{equation}
\dot{A}_{1e}=-(\kappa +\gamma +i\delta +i\Delta )A_{1e}-\sqrt{2}%
igA_{2g}-i\Omega A_{1e}.
\end{equation}%
Solving Eq. (6) will reveal all the physics. Let the initial state of the
system be $\left\vert 0,g\right\rangle $. Considering the weak limit of the
driving field again, we can get $\bar{A}_{0g}$ $\rightarrow $ $1$, and the
Eq. (6) are closed. Thus, Eq. (6) can be easily solved. In the following, we
will only consider the question in the steady-state case. In addition, the
steady-state solution of Eq. (6) can be analytically obtained, but the
concrete form are quite cumbersome, so we further neglect the high-order
terms in $\Omega $ (weak driving) and obtain
\end{subequations}
\begin{equation}
\bar{A}_{1g}=-\frac{i\Omega (\gamma +i\delta )}{g^{2}+\tilde{\Delta}},
\end{equation}%
\begin{equation}
\bar{A}_{0e}=-\frac{g\Omega }{g^{2}+\tilde{\Delta}},
\end{equation}%
\begin{equation}
\bar{A}_{2g}=-\frac{\Omega ^{2}[g^{2}-(\gamma +i\delta )^{2}-\tilde{\Delta}]%
}{\sqrt{2}(g^{2}+\tilde{\Delta})(g^{2}+(\kappa +i\Delta )^{2}+\tilde{\Delta})%
},
\end{equation}%
\begin{equation}
\bar{A}_{1e}=\frac{ig^{2}\Omega ^{2}(\gamma +\kappa +i\delta +i\Delta )}{%
(g^{2}+\tilde{\Delta})(g^{2}+(\kappa +i\Delta )^{2}+\tilde{\Delta})},
\end{equation}%
with $\tilde{\Delta}=(\gamma +i\delta )(\kappa +i\Delta ).$ Thus Eq. (\ref%
{nn}) can be rewritten as $g^{(2)}(0)=\frac{2p_{2}}{(p_{1}+2p_{2})^{2}},$
with $p_{1}=\left\vert \bar{A}_{1g}\right\vert ^{2},p_{2}=\left\vert \bar{A}%
_{2g}\right\vert ^{2}$. In the weak-driving case, we can easily get $%
p_{1}\gg $ $p_{2},$ then the second-order correlation function
\begin{equation}
g^{(2)}(0)=\frac{\left\vert g^{2}+\gamma +i\delta \right\vert ^{2}\left\vert
g^{2}-(\gamma +i\delta )^{2}-\tilde{\Delta}\right\vert ^{2}}{\left\vert
\gamma +i\delta \right\vert ^{4}\left\vert g^{2}+(\kappa +i\Delta )^{2}+%
\tilde{\Delta}\right\vert ^{2}},  \label{aa}
\end{equation}%
\begin{equation}
\bar{n}=\frac{\Omega ^{2}\left\vert \gamma +i\delta \right\vert ^{2}}{%
\left\vert g^{2}+\tilde{\Delta}\right\vert ^{2}}.
\end{equation}

In order to show the validity of the above results, we use the Markovian
master equation for the model, that is,
\begin{equation}
\dot{\rho}=-i[H,\rho ]+\kappa L[a]\rho +\gamma L[\sigma ^{-}]\rho
\label{mmm}
\end{equation}%
where $H$ is the Hamiltonian given by (\ref{a}), $\rho $ is the density
operator of the whole composite system, and $L[\hat{d}]\rho =2\hat{d}\rho
\hat{d}^{\dagger }-\hat{d}^{\dagger }\hat{d}\rho -\rho \hat{d}^{\dagger }%
\hat{d},(\hat{d}=\hat{a},\hat{\sigma}^{-})$ is the dissipator. In addition,
we don't consider the thermal photons for simplicity \cite{xuxunwei}. Since
the steady-state solution is needed for our purpose, we will directly employ
a numerical way to solving Eq. (\ref{mmm}) for the steady state $\rho _{s}$
\cite{tan1}. So the second-order correlation function can be directly
obtained by $g^{(2)}(0)=\frac{Tr[\rho _{s}a^{\dagger 2}a^{2}]}{[Tr(\rho
_{s}a^{\dagger }a)]^{2}},$ and the mean photon number is given by $\bar{n}=$
$Tr(\rho _{s}a^{\dagger }a)$.

In what follows, we will employ both the analytical method given by Eq. (\ref%
{aa}) and the numerical way as Eq. (\ref{mmm}) to study the properties of
the equal-time correlation function $g^{(2)}(0)$. As is shown in Fig. 1(d),
we plot $g^{(2)}(0)$ as a function of the cavity-laser detuning $\Delta ,$ a
different story takes place with resonant condition when the model under the
strong coupling region; the detuning $\Delta /\kappa =-50(25)$ corresponds
to the red lines in Fig. 1(b), which means the transition $\left\vert
0\right\rangle $ $\rightarrow \left\vert 2_{\pm }\right\rangle $ and $%
g^{(2)}(0)\gg 1,$ indicating that the photon bunching for the cavity. In
addition, when $\Delta /\kappa =-50,$it also indicates a quasi-Dark state ($%
\left\vert \tilde{d}\right\rangle \propto g$ $\left\vert 0,g\right\rangle
-\Omega \left\vert 0,e\right\rangle $), which exhibit the strong bunching
behavior. At the point of $\Delta /\kappa =-25(1\pm \sqrt{5}),$ $%
g^{(2)}(0)\ll 1$, indicating the complete photon blockade due to the
suppressed two-photon process, corresponds to the blue lines in Fig. 1(b).
When $\Delta =0,$ the system does not exhibit the strong bunching behavior
as the bare cavity \cite{coherent}, that means the absorption of the first
photon can not enhance the absorption of the subsequent photons and the
expected photon-induced tunnelling phenomenon disappeared owing the
cavity-atom off-resonance.

\section{ Atomic center-of-mass motion on photon statistics}

\subsection{ The model and equal-time correlation}

In this section, we consider the effect of atomic center-of-mass motion on
the photon statistics. As shown in Fig. 2(a), a two-level system which is
confined by a harmonic potential with the trap frequency $\nu $. In this
configuration, the two-level system and cavity are same as before. We
suppose that the model is a one-dimensional model and restrict the motion of
the atom along the $x$ axis. So the Hamiltonian of this system under the
dipole approximation reads \cite{Phase Space}
\begin{figure}[tbp]
\centering
\includegraphics[width=0.65\columnwidth,height=3in]{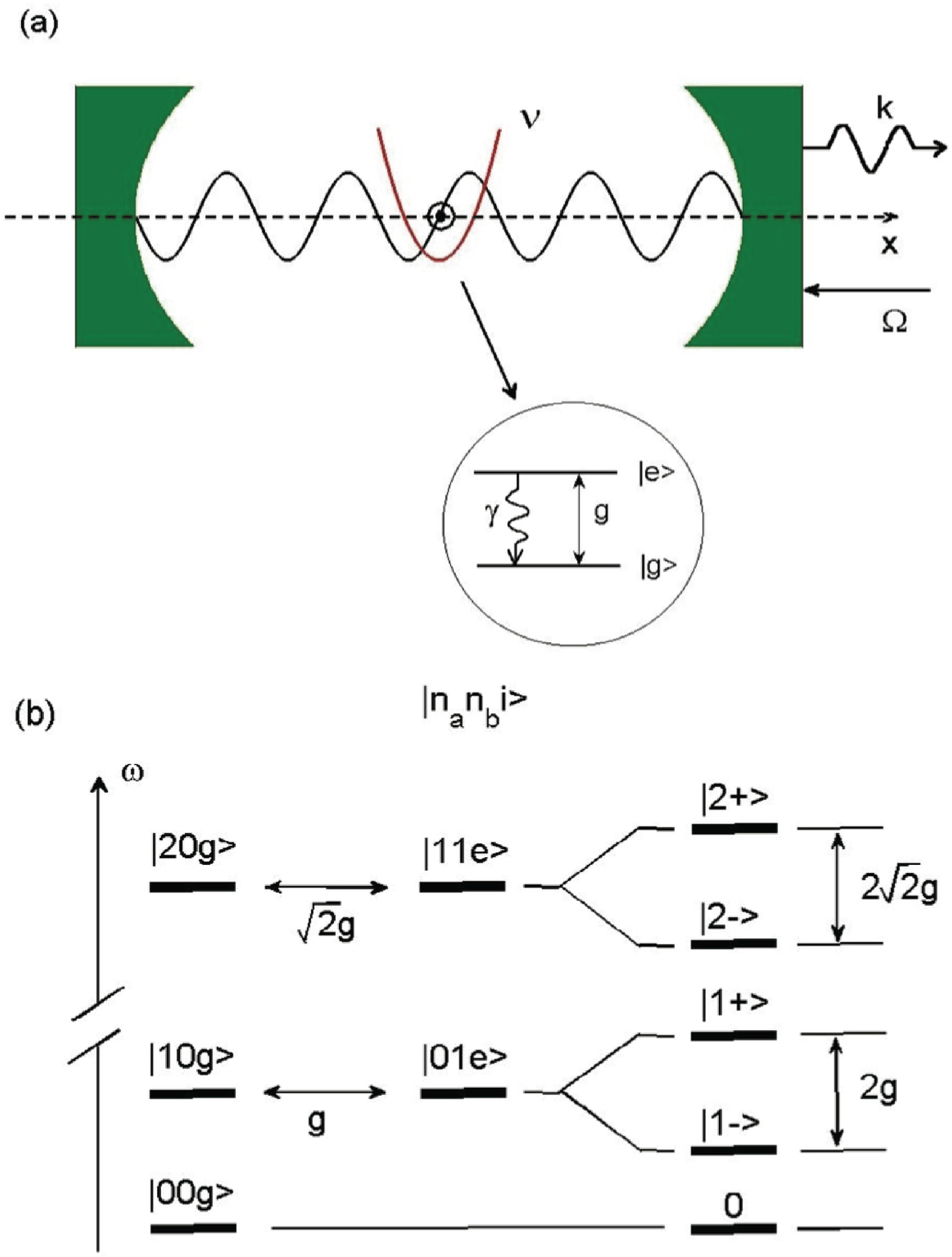} \hspace*{%
-0.5cm} \includegraphics[width=0.7\columnwidth,height=1.5in]{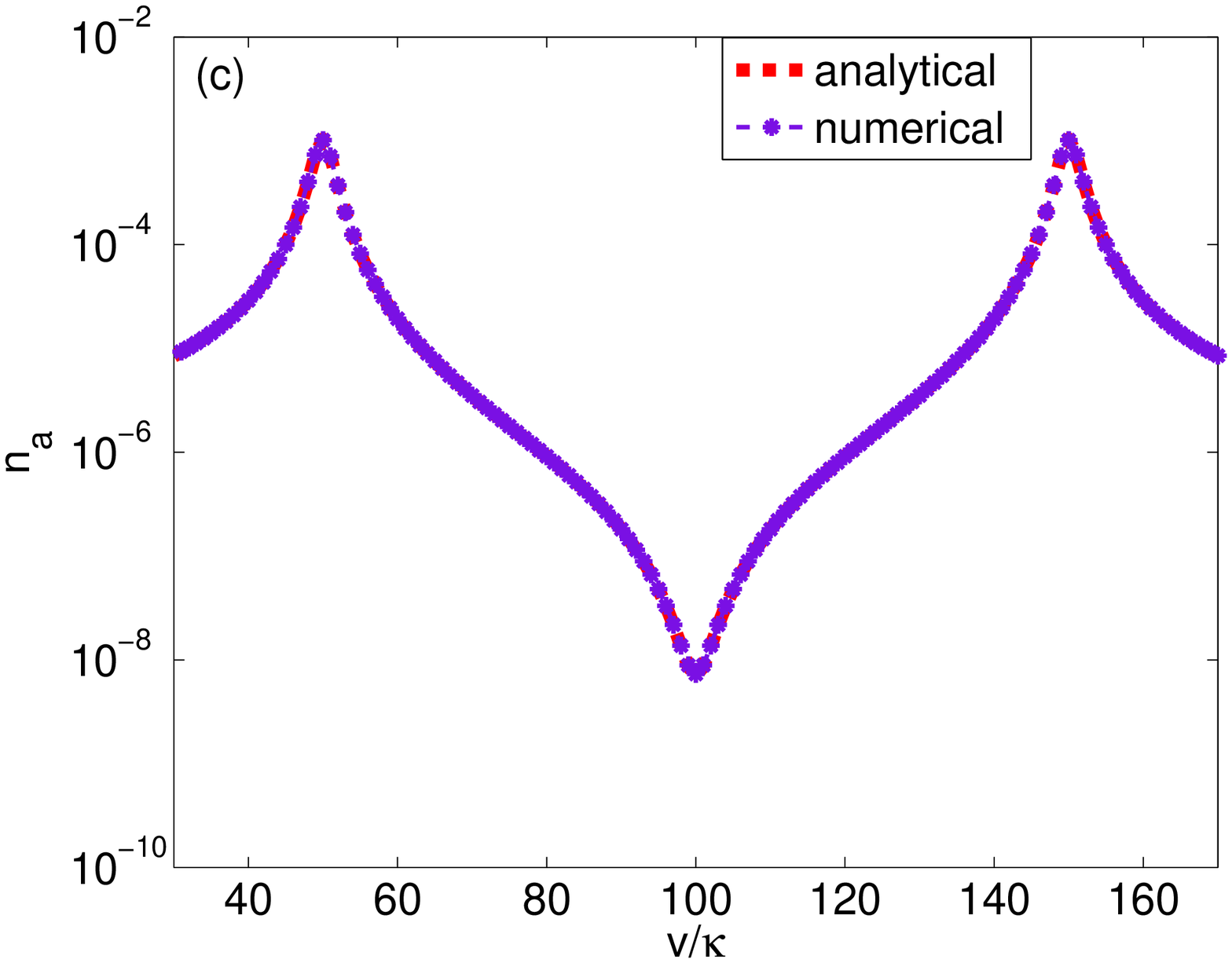} %
\includegraphics[width=0.7\columnwidth,height=1.5in]{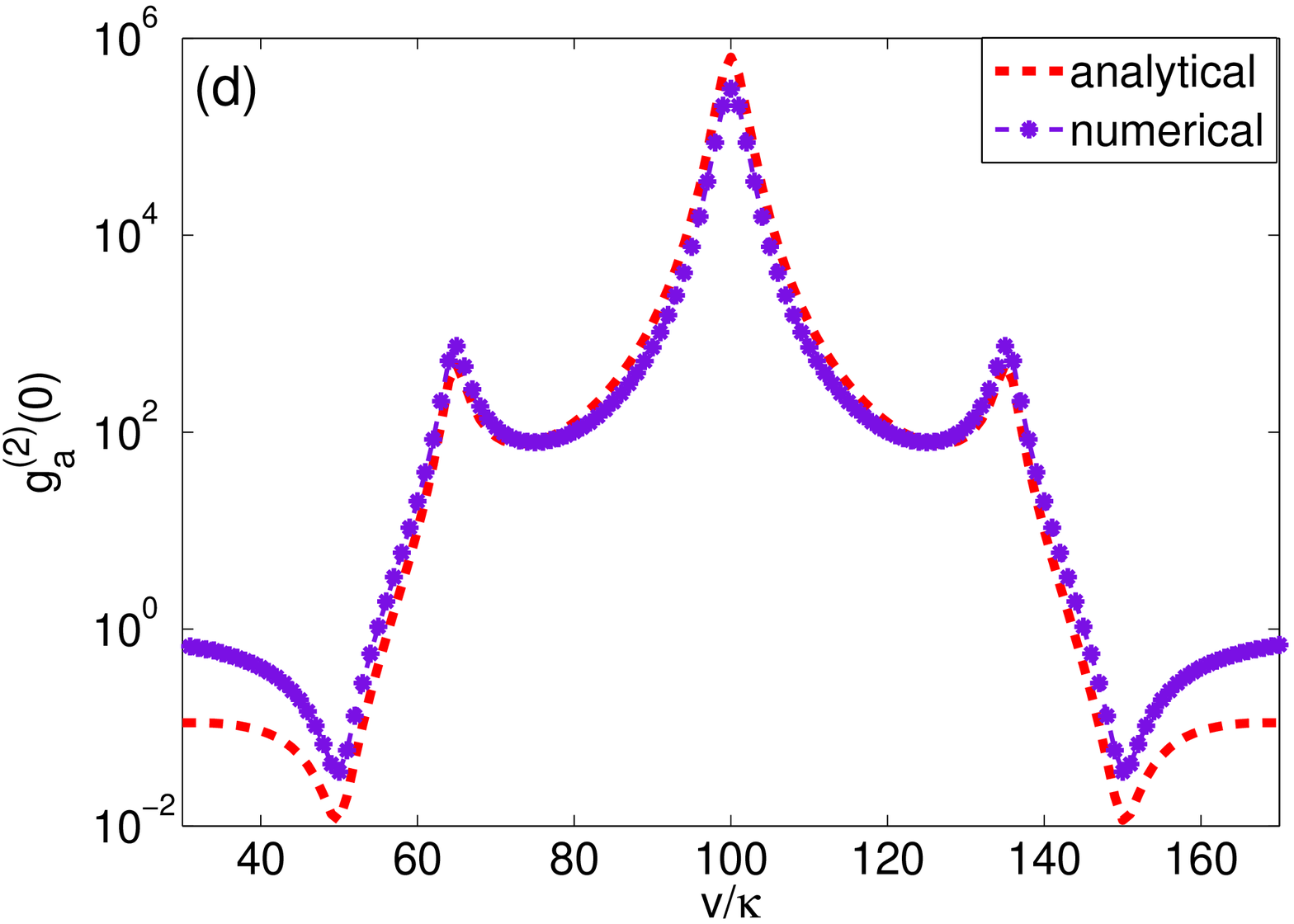}
\caption{(color online). (a).The center-of-mass motion of a trapped atom
under the assumption that the atom is confined within a high-finesse cavity
which is pumped by a laser. (b) Diagram of the eigensystem of the
Hamiltonian under the resonance condition and level diagram is limited in
the subspace spanned by the zero-, one-, and two-photon states. States are
labeled by $\left\vert n_{a},n_{b},i\right\rangle $ denoting the photon
numbers of cavity mode , phonon number of mechanical mode and the level of
atom. (c),(d): The mean photon numbers of the cavity and the equal-time
second-order function $g^{(2)}(0)$ vs. the trap frequency $\protect\nu $
respectively. The red curves are approximately analytical solution of Eq.
(30) and Eq. (31) while the purple curves are numerical results based on the
solution of the quantum master equation Eq. (32). The parameters are chosen
the same as Fig. 1. }
\end{figure}
\begin{align}
H& =\omega _{a}a^{\dag }a+\omega _{e}\sigma ^{+}\sigma ^{-}+\frac{\hat{p}^{2}%
}{2m}+\frac{m\upsilon ^{2}\hat{x}^{2}}{2}  \label{1} \\
& +\tilde{g}(k\hat{x})(\sigma ^{+}a+a^{\dag }\sigma ^{-})+\Omega \left(
a^{\dag }e^{-i\omega _{L}t}+ae^{i\omega _{L}t}\right) ,  \notag
\end{align}%
where $\tilde{g}$ is the position-dependent coupling coefficient, $\Omega $
and $\omega _{L}$ are related to the power and frequency of the driving
laser, respectively, $k$ denotes the wave number of the field. Note that the
third and forth terms represent the kinetic energy and the harmonic
potential respectively, and $m$ represents the mass of the atom. The
strength of the interaction of the two-level system with the single mode of
cavity is characterized by the coupling operator which is given by \cite%
{Marc}
\begin{equation}
\tilde{g}(k\hat{x})=\tilde{g}\cos (kx\cos \phi +\varphi ),
\end{equation}%
where $\phi $ is the angle between wave vector and axis of the motion, $%
\varphi $ accounts for the displacement of the trap center with respect to
the origin. For convenience we set $\phi =0,$ $\varphi =\frac{\pi }{2}$ ,
and we also introduce the annihilation and creation operators $b$ and $%
b^{^{\dag }}$ of a quantum of vibrational energy, therefore, the position
and canonically-conjugated momentum of the atom are given by $\hat{x}=\sqrt{%
\frac{\hbar }{2m\upsilon }}(b+b^{^{\dag }}),$ $\hat{p}=i\sqrt{\frac{\hbar
m\upsilon }{2}}(b^{^{\dag }}-b)$. In addition, the third and forth terms of
Hamiltonian (\ref{1}) can be rewritten as $\hbar \upsilon (b^{\dag }b+\frac{1%
}{2})$. Omitting the constants, Hamiltonian (\ref{1}) becomes (we set $\hbar
=1$ hereafter)
\begin{eqnarray}
H_{0} &=&\omega _{a}a^{\dag }a+\omega _{e}\sigma ^{+}\sigma ^{-}+\upsilon
b^{\dag }b+g(a^{\dag }\sigma ^{-}+a\sigma ^{+})\left( b^{\dag }+b\right) +
\notag \\
&&\Omega \left( a^{\dag }e^{-i\omega _{L}t}+ae^{i\omega _{L}t}\right) ,
\end{eqnarray}%
with $g=$ $\tilde{g}\frac{\omega _{c}}{c}\sqrt{\frac{\hbar }{2m\upsilon }}$.
In the frame rotated at the laser frequency $\omega _{L},$ we obtain
\begin{eqnarray}
H^{\prime } &=&\Delta a^{\dag }a+\delta \sigma ^{+}\sigma ^{-}+\upsilon
b^{\dag }b+g(a^{\dag }\sigma ^{-}+a\sigma ^{+})\left( b^{\dag }+b\right)
\notag \\
&&+\Omega \left( a^{\dag }+a\right) ,
\end{eqnarray}

Now, we focus on the three-mode resonant interaction in which the cavity and
atom exchange a photon by absorbing or emitting a phonon in the mechanical
mode. We set $\Delta =\delta +\upsilon $ and assume $\left\vert \delta
\right\vert \gg \frac{g}{2}$. In order to study the dynamics of the system,
we would like to turn to a rotation framework subject to the transformation $%
\hat{u}(t)=\exp [-i\hat{R}t]$ with $R=\delta (\sigma ^{+}\sigma ^{-}-b^{\dag
}b)$. With the rotating-wave approximation at large detuning $\delta $, the
effective Hamiltonian can be given by
\begin{align}
H_{\mathrm{eff}}& =(\delta +\upsilon )(a^{\dag }a+b^{\dag }b)+g(a^{\dag
}b\sigma ^{-}+ab^{\dag }\sigma ^{+})  \notag \\
& +\Omega \left( a^{\dag }+a\right) .  \label{3}
\end{align}%
The second term in the first line of Eq. (\ref{3}) describes the effective
nonlinear coupling proportional to coupling strength $g$ which describes the
coherent photon-phonon exchange between the cavity mode and mechanical mode
mediated by atomic absorption or emission of a photon. The effective
Hamiltonian can be diagonalized in the absence of a driving \cite{fliuds}.
The eigenstates distinguished by different numbers of photons and phonons
can be expressed as
\begin{equation}
\left\vert 0\right\rangle =\left\vert 0,0,g\right\rangle ,
\end{equation}%
\begin{equation}
\left\vert 1_{\pm }\right\rangle =\frac{1}{\sqrt{2}}(\left\vert
1,0,g\right\rangle \pm \left\vert 0,1,e\right\rangle ,
\end{equation}%
\begin{equation}
\left\vert 2_{\pm }\right\rangle =\frac{1}{\sqrt{2}}(\left\vert
2,0,g\right\rangle \pm \left\vert 1,1,e\right\rangle ,
\end{equation}%
where $\left\vert n_{a},n_{b},i\right\rangle (i=e,g)$ are the preferential
basis with $n_{a},n_{b}$ denoting the photon number and phonon number,
respectively, and $\left\vert i\right\rangle (\left\vert g\right\rangle
,\left\vert e\right\rangle )$ represents the ground and the excited states
of the atom. The low-energy level diagram is sketched in Fig. 2 (b), from
which one can see that the states $\left\vert 1,0,g\right\rangle $ and $%
\left\vert 0,1,e\right\rangle $ are superposed to form two eigenstates $%
\left\vert 1_{\pm }\right\rangle $ splitted by $2g$, and the states are $%
\left\vert 2,0,g\right\rangle $ and $\left\vert 1,1,e\right\rangle $ are
superposed to form another two eigenstates $\left\vert 2_{\pm }\right\rangle
$ splitted by $2\sqrt{2}g$. Intuitively, one can find, from the level
diagram, that the resonant absorption of a photon with frequency $\omega
_{a}\pm g$ to reach the state $\left\vert 1_{\pm }\right\rangle $ `block'
absorb the second photon with the same frequency because of the detuning
from the other levels \cite{Werner,Rabl,Nunnenkamp,Ferretti1}.

Using the same approach, we can obtain the following results for the atomic
center-of-mass motion system. So we can also assume that the state of the
composite system is
\begin{eqnarray}
\left\vert \Psi \right\rangle _{com} &=&A_{00}\left\vert 0,0,g\right\rangle
+A_{10}(\left\vert 1,0,g\right\rangle +A_{01}\left\vert 0,1,e\right\rangle
\label{phicom} \\
&&+A_{11}\left\vert 1,1,e\right\rangle +A_{20}\left\vert 2,0,g\right\rangle
\end{eqnarray}
The equations about the amplitudes in Eq.(\ref{phicom}) as follows.
\begin{subequations}
\begin{equation}
\dot{A}_{10}=-(\kappa +i\Delta )A_{10}-i\Omega A_{00}-igA_{01}-\sqrt{2}%
i\Omega A_{20},
\end{equation}%
\begin{equation}
\dot{A}_{01}=-(\gamma +\Gamma +i\Delta )A_{01}-igA_{10}-i\Omega A_{11},
\end{equation}%
\begin{equation}
\dot{A}_{20}=-2(\kappa +i\Delta )A_{20}-\sqrt{2}igA_{11},
\end{equation}%
\begin{equation}
\dot{A}_{11}=-(\kappa +\gamma +\Gamma +2i\Delta )A_{11}-\sqrt{2}%
igA_{20}-i\Omega A_{11},
\end{equation}%
\begin{figure}[tbp]
\centering
\includegraphics[width=0.75\columnwidth,height=2.35in]{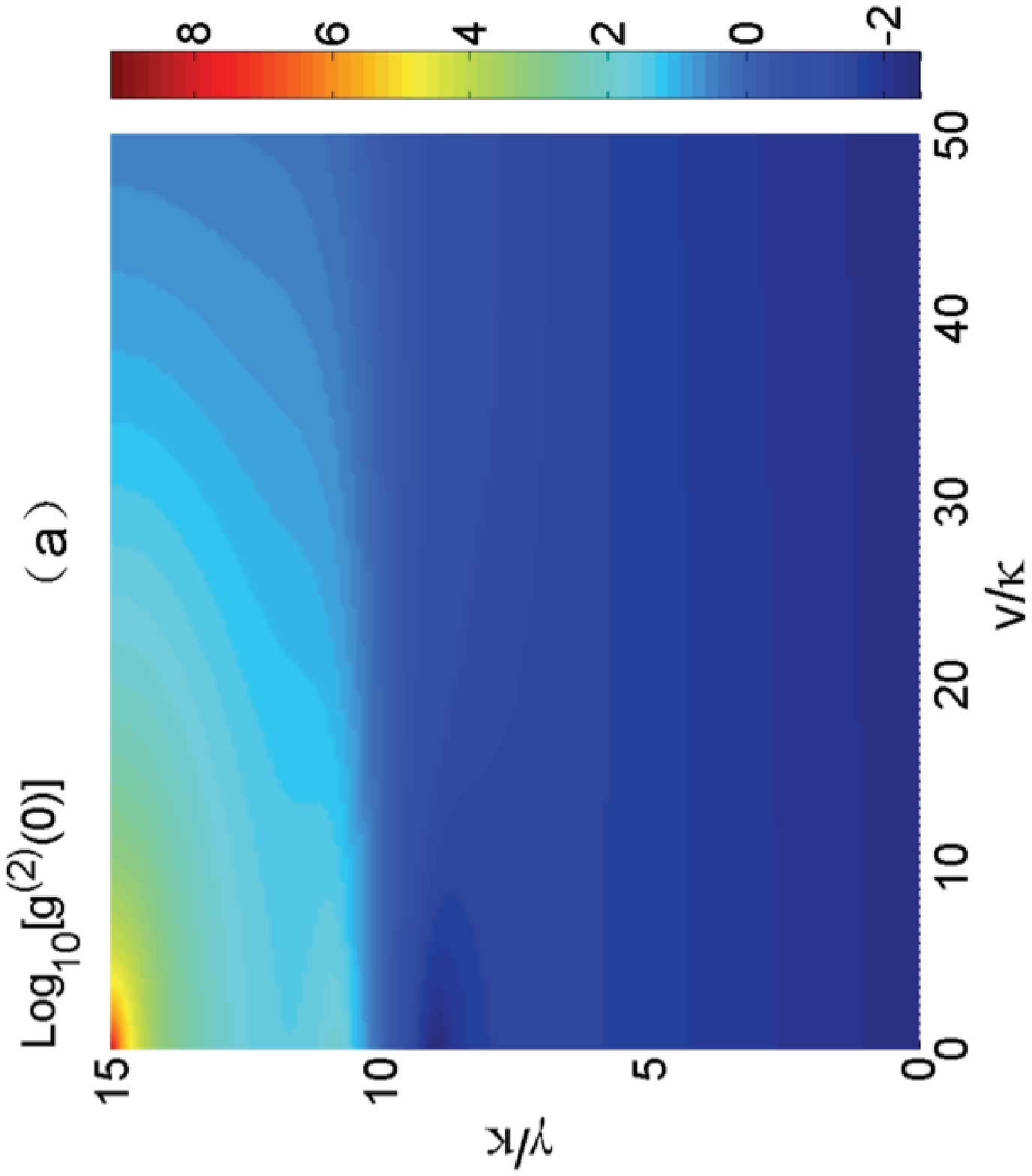}
\centering
\includegraphics[width=0.75\columnwidth,height=2.35in]{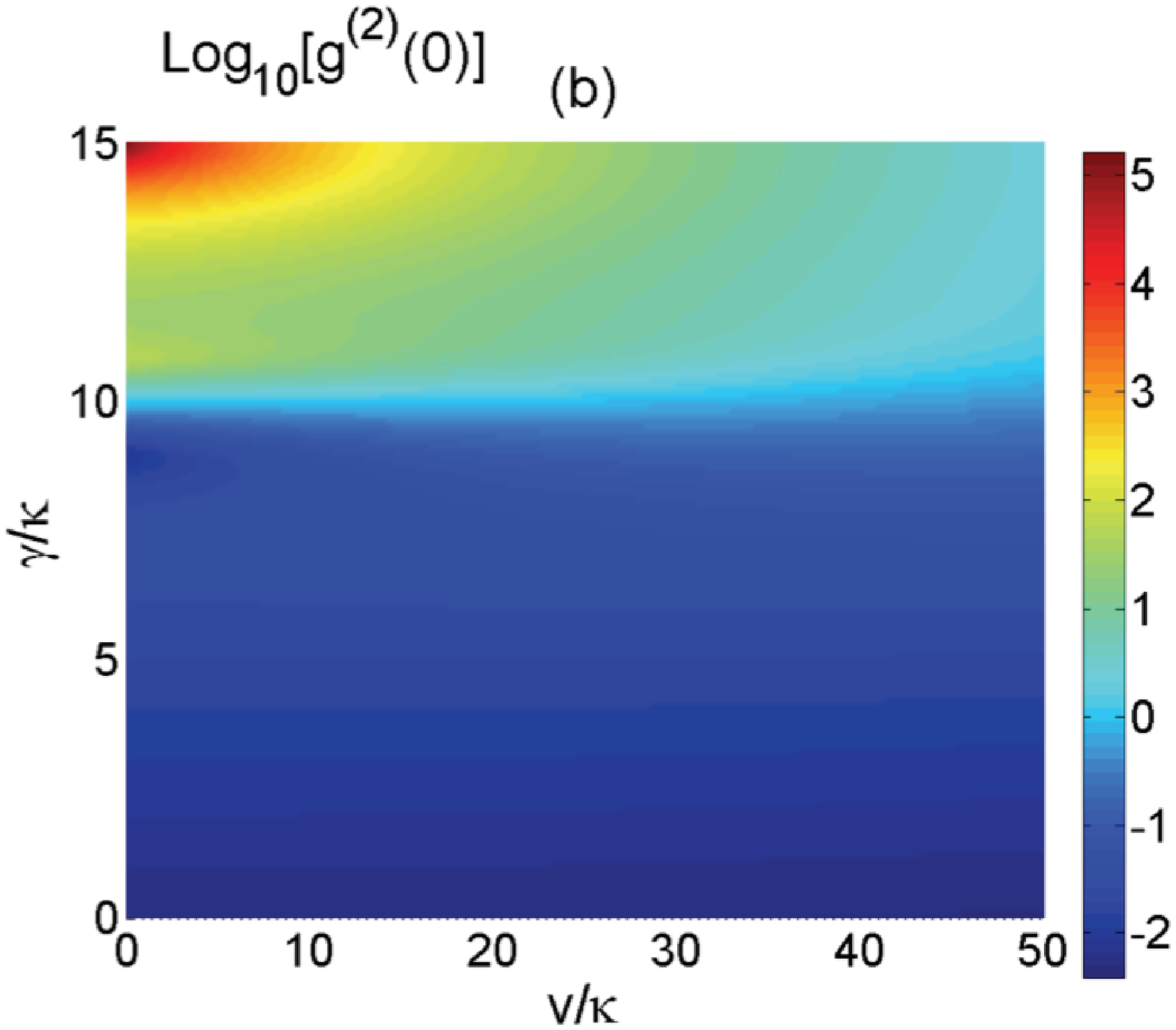}
\centering
\includegraphics[width=0.6\columnwidth,height=2in]{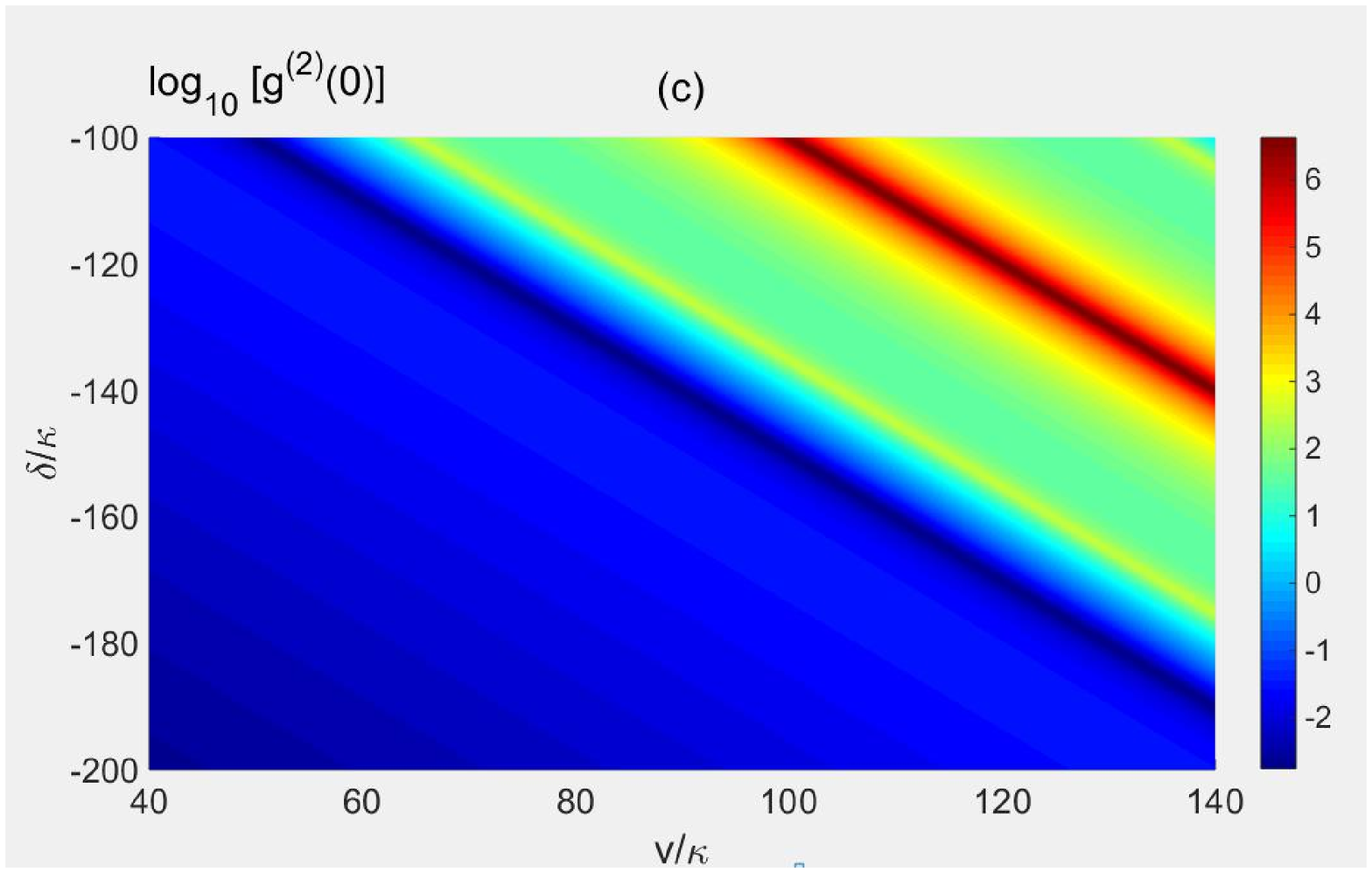} \centering
\caption{(color online).We plot the $\log _{10}\left[ g^{(2)}(0)\right] $ as
a function of $\protect\nu $ and $\protect\gamma $ for (a) $\Gamma /\protect%
\kappa =0.1$ and for (b) $\Gamma /\protect\kappa =1$. The other parameters
are $\protect\delta /\protect\kappa =-50$, $\Omega /\protect\kappa =0.1$.
For (c), we plot the $\log _{10}\left[ g^{(2)}(0)\right] $ as a function of $%
\protect\nu $ and $\protect\delta $, we take $\protect\gamma /\protect\kappa %
=1,$ $\Gamma /\protect\kappa =0.1$, $\Delta=\protect\delta+\protect\nu,$%
 other parameters are the same as (a) and (b). }
\end{figure}

Solving Eq. (24) will reveal all the physics. Let the initial state of the
system be $\left\vert 0,0,g\right\rangle $. Considering the weak limit of
the driving field again, we can get $\bar{A}_{00}$ $\rightarrow $ $1$, and
obtain the following coupled set of equations for the steady state:

\end{subequations}
\begin{subequations}
\begin{equation}
-(\kappa +i\Delta )\bar{A}_{10}-i\Omega \bar{A}_{00}-ig\bar{A}_{01}-\sqrt{2}%
i\Omega \bar{A}_{20}=0,
\end{equation}%
\begin{equation}
-(\gamma +\Gamma +i\Delta )\bar{A}_{01}-ig\bar{A}_{10}-i\Omega \bar{A}%
_{11}=0,
\end{equation}%
\begin{equation}
-2(\kappa +i\Delta )\bar{A}_{20}-\sqrt{2}ig\bar{A}_{11}=0,
\end{equation}%
\begin{equation}
-(\kappa +\gamma +\Gamma +2i\Delta )\bar{A}_{11}-\sqrt{2}ig\bar{A}%
_{20}-i\Omega \bar{A}_{11}=0,
\end{equation}

then we can obtain
\end{subequations}
\begin{equation}
\bar{A}_{10}=-\frac{i\Omega \tilde{\gamma}}{g^{2}+\tilde{\gamma}\tilde{\kappa%
}},
\end{equation}%
\begin{equation}
\bar{A}_{01}=-\frac{g\Omega }{g^{2}+\tilde{\gamma}\tilde{\kappa}},
\end{equation}%
\begin{equation}
\bar{A}_{11}=-\frac{ig\tilde{\kappa}\Omega ^{2}}{(g^{2}+\tilde{\gamma}\tilde{%
\kappa})(g^{2}+\tilde{\kappa}^{2}+\tilde{\gamma}\tilde{\kappa})},
\end{equation}%
\begin{equation}
\bar{A}_{20}=\frac{g^{2}\Omega ^{2}}{\sqrt{2}(g^{2}+\tilde{\gamma}\tilde{%
\kappa})(g^{2}+\tilde{\kappa}^{2}+\tilde{\gamma}\tilde{\kappa})},
\end{equation}%
where $\tilde{\gamma}=\gamma +\Gamma +i(\delta +\upsilon ),$ $\tilde{\kappa}%
= $ $\kappa +i(\delta +\upsilon ).$ Thus with $p_{1}=\left\vert \bar{A}%
_{10}\right\vert ^{2},p_{2}=\left\vert \bar{A}_{20}\right\vert ^{2}$, $%
p_{1}\gg $ $p_{2},$ the second-order correlation function becomes
\begin{equation}
\ g_{a}^{(2)}(0)\approx \frac{2p_{2}}{p_{1}^{2}}=\frac{g^{4}(g^{2}+\tilde{%
\gamma}^{\ast }\tilde{\kappa}^{\ast })(g^{2}+\tilde{\gamma}\tilde{\kappa})}{%
(g^{2}+\tilde{\kappa}^{\ast 2}+\tilde{\gamma}^{\ast }\tilde{\kappa}^{\ast
})(g^{2}+\tilde{\kappa}^{2}+\tilde{\gamma}\tilde{\kappa})|\tilde{\gamma}|^4},  \label{bb}
\end{equation}%
and the mean photon number is
\begin{equation}
n_{a}\approx p_{1}=\frac{\Omega ^{2}\left\vert \tilde{\gamma}\right\vert ^{2}%
}{(g^{2}+\tilde{\gamma}^{\ast }\tilde{\kappa}^{\ast })(g^{2}+\tilde{\gamma}%
\tilde{\kappa})}.
\end{equation}

Meanwhile the master equation is
\begin{equation}
\dot{\rho}=-i[H_{\mathrm{eff}},\rho ]+\kappa L[a]\rho +\gamma L[\sigma
^{-}]\rho +\Gamma L[b]\rho ,  \label{n}
\end{equation}%
where $H_{\mathrm{eff}}$ is the effective Hamiltonian given by (\ref{3}),
and $L[\hat{d}]\rho =2\hat{d}\rho \hat{d}^{\dagger }-\hat{d}^{\dagger }\hat{d%
}\rho -\rho \hat{d}^{\dagger }\hat{d},(\hat{d}=\hat{a},\hat{b},\hat{\sigma}%
^{-})$ is the dissipator.

In what follows, we will employ both the analytical method given by Eq. (\ref%
{bb}) and the numerical way as Eq. (\ref{n}) to study the properties of the
equal-time correlation function $g^{(2)}(0)$. As is shown in Fig. 2 (c), we
plot $g^{(2)}(0)$ as a function of the trap frequency $\nu $. We find that
the photon statistic properties can be controlled by tuning the trap
frequency $\nu .$ Here, the red curves are plotted using the analytical
expression Eq. (\ref{bb}) while the grape curves are based on the numerical
solution of Eq. (\ref{n}). One can see that at $\nu /\kappa =100(\Delta =0),$
$g^{(2)}(0)\gg 1$ in Fig. 2(d), where the photon satisfy the
super-Poissonian distribution. In other words, it indicates that at this
point photons occur the bunching behavior. This can be explained that the
system presents destructive interference that suppresses the population in $%
\left\vert 1,0,g\right\rangle $. At this point, the system is driven into a
dark state $\left\vert dark\right\rangle \propto g$ $\left\vert
0,0,g\right\rangle -\Omega \left\vert 0,1,e\right\rangle $ which is similar
to the electromagnetically induced transparency \cite{science330,lukin}$.$
In the dark sate, $\left\vert 0,1,e\right\rangle $ remains populated,
allowing transitions to $\left\vert 1,1,e\right\rangle $ which is strongly
coupled to $\left\vert 2,0,g\right\rangle $. The probability of two photons
inside the cavity is resonantly enhanced at this point which corresponds to
the peak in the correlation function of $g^{(2)}(0)$. Moreover, we see that
at $\nu /\kappa =100,$ the strong bunching regime is accompanied by a
suppression of the photon number, the photon-induced tunneling occurs due to
the probability of single-photon emission decreases, while the probability
for photon pair generation increases. This behavior can also be occurred
when $\nu /\kappa =100\pm 25\sqrt{2},$ because of the transition $\left\vert
0\right\rangle $ $\rightarrow \left\vert 2_{+}\right\rangle $ which is a
double photon resonance process. At the frequency $\nu /\kappa =50$ and $150$%
, the detuning $\Delta /\kappa =-50$ and $50$ correspondingly, the
second-order correlation function $g^{(2)}(0)\ll 1$ (i.e., two dips), which
shows the anti-bunching and sub-Poissonian distribution. This is consistent
with our intuitive analysis given above. That is, the frequency of the
driving is resonant with the transition between the state $\left\vert
0\right\rangle $ and $\left\vert \pm 1\right\rangle $, which `blocks' the
absorption of the second photon due to the detuning.

So far, we have demonstrated the quantum properties with the analytical and numerical methods
both in the J-C model and atomic center-of-mass motion model. By comparing
the two curves in Fig. 1(d) , we find that by introduced phonon degree of
freedom, the photon statistics changes as a function of $\Delta $, and
photon blockade effect occurs at the different $\Delta $ compared
with the J-C model which can also be demonstrated by the analytical
expressions of second-order correlation function in Eq. (11) and Eq. (30).
We can also find that the trap frequency $\nu$ leads to an obvious displacement of the maximal peak, whilst it leads to
a new bunching peak. 

To find the effects of the decay and the detuning on the photon blockade, we
plot the logarithm of the correlation function $g^{(2)}(0)$ as a function of $%
\gamma $ and $\nu $ in Fig. 3. Fig. 3 gives a boundary for different photon
distributions such as super-Poissonian and sub-Poissonian distributions. The
two regions are distinguished by $\gamma /\kappa \sim 10$ in both Fig. 3(a)
and 3(b). These two figures show how the second-order correlation function varies with the decay rates. In particular, one can find that the maximum of  $g^{(2)}(0)$  occurs at the top of the figures ($\gamma/\kappa=15$), but the minimum of  $g^{(2)}(0)$  (anti-bunching) is just below $\gamma /\kappa \sim 10$. Between them, there also exists a bunching peak. In Fig. 3 (c),  $g^{(2)}(0)$  between bunching
and anti-bunching at the frequency $\nu $ and detuning $\delta $ is
illustrated. It is apparent that the photon blockade happens in the region $%
\left\vert \Delta /\kappa \right\vert \approx g$, which can be obtained from Eq. (30) by setting the numerator of $g^{(2)}(0)$ vanishes for small dissipations.

\subsection{Delayed coincidence correlation}

In addition to the second-order correlation function for equal time
discussed in the previous section, some quantum signatures can also be
manifested from the photon intensity correlations with the nonzero delay. In
this part, we give a discussion of the evolution of the two-time correlation
function, which is defined for stationary state by \cite{Ferretti,87}
\begin{equation}
g^{(2)}(\tau )=\frac{\left\langle a^{\dagger }(0)a^{\dagger }(\tau )a(\tau
)a(0)\right\rangle }{\langle a^{\dagger }a\rangle ^{2}},
\end{equation}%
Eq. (33) can be rewritten, based on the correlations of classical light
intensity $I$, as $g^{(2)}(\tau )=\frac{\left\langle I(\tau
)I(0)\right\rangle }{\left\langle I(0)\right\rangle ^{2}}$. With the Schwarz
inequality for random variables:
\begin{equation}
\left\langle I_{1}I_{2}\right\rangle ^{2}\leq \left\langle
I_{1}^{2}\right\rangle \left\langle I_{2}^{2}\right\rangle,
\end{equation}
which means that
\begin{equation}
\left\langle I(\tau )I(0)\right\rangle ^{2}\leq \left\langle I(\tau
)^{2}\right\rangle \left\langle I(0)^{2}\right\rangle .
\end{equation} The inequality is saturated if $\tau=0$.
Thus one can further find that \cite{scully,xiao}%
\begin{equation}
g^{(2)}(\tau )\leq g^{(2)}(0).  \label{inequality}
\end{equation}%
\begin{figure}[tbp]
\centering
\includegraphics[width=1\columnwidth,height=2in]{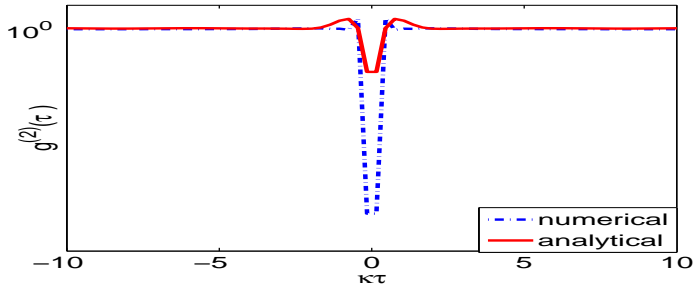} \centering
\caption{(color online).The evolution of finite time delay correlation
function $\ g^{(2)}(\protect\tau )$ of the atomic center-of-mass motion
cavity mode. The red line indicate the analytical expression and the blue
dash line denotes the numerical simulation. The parameters are chosen as $g/%
\protect\kappa =20$ , $\protect\delta /\protect\kappa =-70$, $\Omega /%
\protect\kappa =4$, $\protect\gamma /\protect\kappa =4,$ $\Gamma /\protect%
\kappa =0.1$, $\protect\nu /\protect\kappa =50$ . }
\end{figure}
Similar to the classical inequality, $g^{(2)}(0)>1$ at zero delay, and
violation of the inequality at finite delay is an unique signature of
quantum system. Next we will calculate the delayed second-order correlation
function for the cavity mode. The two-time correlation functions can be
understood in terms of the five-level model discussed above. For this
purpose, we may use the standard method to calculate the two-time
correlation function with a conditional state as the initial condition \cite%
{quantum noise}. The unnormalized state after the annihilation of a photon
in the cavity is $a\left\vert \Psi \right\rangle =\bar{A}_{10}(\left\vert
0,0,g\right\rangle +\bar{A}_{11}\left\vert 0,1,e\right\rangle +\sqrt{2}\bar{A%
}_{20}\left\vert 1,0,g\right\rangle $. By solving the Eq. (11)(a)-(d) with
this state as the initial condition, one can obtain the time-delayed
correlation function as
\begin{equation}
g^{(2)}(\tau )=\frac{\left\vert A_{10}(\tau )\right\vert ^{2}}{\left\vert
\bar{A}_{10}\right\vert ^{4}}.
\end{equation}%
The positive derivative of the time-delayed correlation function with
respect to variable $\tau $ represents a nonclassical behavior of the
system. Additionally, it is important for the experimental observation of
the anti-bunching, since the precise measurement of the equal-time
correlation function may be challenging \cite{tunable}. Both the analytical
and numerical results of the two types cavity mode are plotted in Fig. 4. We
can observe that according to the system nonlinearity the time-dependent
second-order correlation function shows the quantum signature which violates
the inequality in Eq. (\ref{inequality}) because $g^{(2)}(\tau )$ going
beyond their initial value at finite time delay. We note that the behavior
of the second-order correlation function in this case is similar to the
single-photon diode in the semiconductor microcavities coupled via $\chi (2)$
nonlinearities \cite{shen}. This can be attributed to the fact the strong
nonlinearity of our system for both the zero and the finite time delay
correlation functions.

\section{Conclusion}

We have studied the photon blockade effect in CQED system weakly driven by a
monochromatic laser field, in which an atom is trapped inside. The spirit of
this scheme is to effectively take advantage of the atomic center-of-mass
motion instead of avoiding it. By the numerical and the approximate
analytical expressions of the one-time and two-time second-order correlation
function for the cavity photons, we have identified several different
processes that can lead to photon-induced tunnelling and photon blockade,
our study provides another way on photon control using a atomic
center-of-mass motion system. In addition, the trap-frequency dependent
blockade effect directly shows the influence of imperfect cooling, whilst it
may induce the other potential applications of atomic center of mass motion
in quantum information processing.

Finally, we would like to say that trapped ions or atoms are well suited for
this purpose as the quantum technology for controlling their degrees of
freedom that is already well developed \cite{Leibfried,Blatt}. The coupling
to the cavity has been successfully employed for implementation of a
photon-mediated entanglement distribution \cite{Leibfried,Blatt,com} and the
cooling the motion of an atom which is trapped by a harmonic trap \cite%
{Marc,Boca,Reimann,Leibrandt}. The combination with the coupling to the
field of high-finesse resonators can bring novel perspective for the
manipulation of the atomic motion.

\section*{Acknowledgement}

This work was supported by the National Natural Science Foundation of China,
under Grant No.11375036 and 11175033, the Xinghai Scholar Cultivation
Plan and the Fundamental Research Funds for the Central Universities under Grant No. DUT15LK35.

\end{document}